\documentclass[english,aps, manuscript]{revtex4}
\usepackage[T1]{fontenc}
\usepackage[latin1]{inputenc}
\usepackage{amsmath}

\makeatletter
\usepackage{babel}
\makeatother
\begin{document}

\title{The Scales of Brane Nucleation Processes}

\author{S. P. de Alwis\protect{}}

\affiliation{Perimeter Institute, 31 Caroline Street N., Waterloo, ON N2L 2Y5,
Canada}

\affiliation{Department of Physics, University of Colorado, Box 390, Boulder,
CO 80309.\\
\\
 \texttt{e-mail: dealwis@pizero.colorado.edu} PACS : 11.25. -w, 98.80.-k}

\begin{abstract}
The scales associated with Brown-Teitelboim-Bousso-Polchinski processes
of brane nucleation, which result in changes of the flux parameters
and the number of D-branes, are discussed in the context of type IIB
models with all moduli stabilized. It is argued that such processes
are unlikely to be described by effective field theory. 
\end{abstract}
\maketitle
\vfill

\eject

\section{Introduction}

There are two approaches to dynamics on the landscape of classical
solutions to (the supergravity limit of) string theory.

1. The universe is the result of a tunneling event as in traditional
quantum cosmology \cite{Hartle:1983ai}\cite{Vilenkin:1983xq}\cite{Linde:1983cm}
except that instead of tunneling from nothing, the pre-tunneling state
is the result of spontaneous compactifiction and decay of high string
states (which may be modeled by a thermal state at a temperature somewhat
below the Hagedorn temperature). The different end states of this
tunneling process are the different four dimensional effective field
theories that are supposed to constitute the landscape of string theory.

2. The second is a process of brane nucleation a la Brown and Teitelboim
(BT) \cite{Brown:1988kg}. This is analogous to the Schwinger effect
of pair production of charged particles by a strong electric field
in two dimensions. The currently observed universe (with a tiny cosmological
constant (CC)) is supposed to be the end result of this process of
brane nucleation. In the generalization of this argument by Bousso
and Polchinski (BP) \cite{Bousso:2000xa} to string theory, this process
can result in getting a universe with a tiny cosmological constant
(CC) even though the brane tension is not parametrically below the
string scale.

In \cite{Brustein:2005yn} Brustein and the author addressed the first
scenario %
\footnote{For related earlier work see \cite{Sarangi:2005cs}.%
}. In this paper we will discuss the second, using type IIB models
with flux compactifications %
\footnote{As for instance in \cite{Giddings:2001yu} - for a complete list of
references see for example \cite{Grana:2005jc}.%
} and non-perturbative terms \cite{Kachru:2003aw} (KKLT) %
\footnote{The analog of the BT and BP processes in this context was discussed
by \cite{Kachru:2002gs}\cite{Frey:2003dm}\cite{Pilo:2004mg}.%
}. In particular we will determine under what conditions this process
could be described by an effective supergravity. This means that given
a state which has such a description, which we will take to be a final
state of the nucelation process such as the observed universe, we
wish to investigate whether its immediate antecedent state is also
describable by effective supergravity. What we will find is that although
generically the antecedent state is strongly curved (at higher than
the string scale) there are non-generic situations where a supergravity
description of the antecedent state is possible. However even in those
situations the required domain wall brane has a tension which is above
the string scale so that it does not seem possible to describe its
nucleation in low energy effective supergravity.

In section 2 we discuss the relation between different scales and
estimate the tension of the relevant brane, and in section 3 estimate
the cosmological constants in the two states and discuss the transition.

\section{A matter of scales - the domain wall tension}

In this section we analyze the tensions of branes and the change in
the CC in the Einstein frame defined by GKP \cite{Giddings:2001yu}.
Here in addition to warping we also put in the dependence of the metric
on the volume modulus.

The metric ansatz is %
\footnote{This factorized ansatz is not suitable for the derivation of a potential
when there is non-trivial warping as pointed out in \cite{deAlwis:2003sn,deAlwis:2004qh}
however it is adequate for estimating the relevant scales provided
we take into acount the effective depedence of the warp factor on
the volume modulus as discussed in \cite{Giddings:2005ff}.%
}:\begin{equation}
ds^{2}=g_{MN}^{S}dx^{M}dx^{N}=e^{\phi/2}[e^{-6u(x)+2A(y)}g_{\mu\nu}(x)dx^{\mu}dx^{\nu}+e^{2u(x)-2A(y)}g_{mn}(y)dy^{m}dy^{n}]\label{metric}\end{equation}
In the above $g_{MN}$ is the ten-dimensional string metric and the
ansatz ensures that $g_{\mu\nu}(x)$ is the four dimensional Einstein
metric and $g_{mn}$ is the Ricci-flat metric on the internal space
which is usually taken to be a Calabi-Yau manifold. $u(x)$ is the
volume modulus and we've ignored the other moduli since they do not
enter into the relation between the Planck and string scales. Also
it is convenient to take the effective volume of the internal manifold
to be at the string scale when $u=0$, in other words we put\begin{equation}
\int_{X}d^{6}ye^{-4A(y)}\sqrt{g^{(6)}(y)}=(2\pi)^{6}\alpha'^{3}\label{6vol}\end{equation}
 Reducing the 10 D action using this ansatz gives\begin{eqnarray}
S & = & \frac{1}{(2\pi)^{7}\alpha'^{4}}[\int d^{4}x\int d^{6}y\sqrt{g^{S(10)}(x,y)}e^{-2\phi}g_{s}^{\mu\nu}R_{\mu\nu}^{S}+...]\label{10action}\\
 & = & \frac{1}{(2\pi)^{7}\alpha'^{4}}[\int d^{6}ye^{-4A(y)}g^{(6)}(y)\int d^{4}x\sqrt{g^{(4)}}R^{(4)}(x)+...]\label{4action}\end{eqnarray}
From the second line we can read off the 4D Planck scale (by identifying
the coefficients of the 4D Einstein action as $M_{p}^{2}/2$)

\begin{equation}
M_{P}^{2}=\frac{2}{(2\pi)^{7}\alpha'^{4}}\int_{X}d^{6}ye^{-4A(y)}\sqrt{g^{(6)}(y)}=\frac{2}{2\pi\alpha'}\label{planckmass}\end{equation}
This is of course measured in the 4D Einsein metric and in the same
metric the string scale can be identified by writing the string action
as,

\begin{eqnarray}
I & \sim & \frac{1}{2\pi\alpha'}\int d^{2}\sigma[\frac{1}{2}g_{MN}^{S}\partial X^{M}\bar{\partial}X^{N}+...]\label{stringaction}\\
 & \sim & \frac{1}{2\pi\alpha'}\int d^{2}\sigma[\frac{1}{2}e^{\phi(x)/2}e^{-6u(x)}e^{2A(y)}g_{\mu\nu}\partial X^{\mu}\bar{\partial}X^{\mu}+...]\nonumber \end{eqnarray}

The effective string tension (measured in the 4D Einstein metric)
is actually space-time dependent unless the dilaton and the volume
modulus are stabilized. Henceforth we will use $u,\,\phi$ for the
stabilized values of the volume modulus and the dilaton. The tension
is also dependent on the warp factor and is given by \begin{equation}
T_{s}\equiv M_{s}^{2}=\frac{e^{\phi/2}e^{-6u}e^{2A(y)}}{2\pi\alpha'}.\label{stringtension}\end{equation}
Thus in these units the string scale in any throat region will depend
on how far down the throat it is measured. The ratio of the Planck
scale to the string scale (which is in fact the physically meaningful
quantity) is $ $\begin{equation}
\frac{M_{s}^{2}}{M_{P}^{2}}=\frac{1}{2}e^{\phi/2}e^{-6u}e^{2A(y)}.\label{stringplanck}\end{equation}
This formula illustrates the fact that in highly warped regions (down
some AdS like throat region that approximates the RS1 \cite{Randall:1999ee}
scenario as in GKP) the Planck scale can be much higher than the string
scale \cite{DeWolfe:2002nn}. The same formula illustrates the fact
that this hierarchy of scales can alternatively be obtained by taking
a large volume ($e^{u}\gg1$) compactification.

Now consider two different flux compactifications separated in (3+1)
dimensions by a domain wall which in IIB theory could be a D5 brane
wrapping a three cycle in the internal manifold. We could also consider
an NS5 brane but the results would be very similar except for the
factor of the dilaton which has only a marginal effect in the following. 

In the string metric the tension of the five brane is given by the
formula $T_{5}=e^{-\phi}/(2\pi)^{5}\alpha'^{3}$. The relevant term
in the five brane action is,\begin{equation}
\frac{1}{(2\pi)^{5}\alpha'^{3}}|\int_{A}\Omega|\int d^{3}xe^{-6u(x)+\phi(x)/2}\sqrt{g^{(3)}(x)},\label{fiveaction}\end{equation}
 where $A$ is the three cycle wrapped by the brane and $\Omega$
is the holomorphic three form on $X$. So the effective tension of
the domain wall measured in the 4D Einstein metric is \begin{equation}
T_{dw}=\frac{|z|}{(2\pi)^{2}\alpha'^{3/2}}e^{-6u+\phi/2},\label{dwtension}\end{equation}
where we have introduced the complex structure modulus associated
with the three cycle wrapped by the brane by writing $\int\Omega=(2\pi)^{3}\alpha'^{3/2}z$.
To estimate the Kaluza-Klein (KK) masses consider for instance the
dilaton kinetic term

\begin{eqnarray}
-\int d^{10}xe^{-2\phi}\sqrt{g_{s}^{(10)}}g_{s}^{MN}(R_{MN}-4\partial_{M}\phi\partial_{N}\phi) & \rightarrow & -\int\sqrt{g^{(4)}(x)g^{6)}(y)}\label{dilaction}\\
{}[e^{-4A}g^{\mu\nu}\partial_{\mu}\phi\partial_{\nu}\phi & + & e^{-8u}g^{mn}\partial_{m}\phi\partial_{n}\phi]\nonumber \end{eqnarray}
Thus we get the KK masses to be \begin{equation}
M_{KK}^{2}\sim\frac{<e^{-4A}>_{X}^{-1}}{2\pi\alpha'}e^{-8u},\label{kkmass}\end{equation}
where we have defined $<e^{-4A}>_{X}=\int_{X}e^{-4A}v^{2}/\int_{X}v{}^{2}$
where $v$ is the KK wave function. Since one expects the KK modes
to be localized at the end of the Klebanov-Strassler (KS) throat (as
in \cite{Goldberger:1999uk} for the RS1 case - see also \cite{Frey:2006wv})
this average may be estimated by using the value far down the throat.
In doing so we need to take into acount the dependence of the warp
factor on the volume modulus as discussed in \cite{Giddings:2005ff}.
In other words we need to write (assuming as before that the volume
modulus is fixed at some space time independent value) \begin{equation}
e^{-4A}=1+e^{-4u}h(y).\label{warp}\end{equation}
Here $h(y)$ is defined to give zero when integrated over $X$ with
metric $g_{mn}.$ Far down some warped throat $h$ becomes large and
takes the value $h_{max}=e^{8K/3Mg_{s}}\equiv e^{-4A_{min}}$ with
$K,M,\, K>>M$ a pair of flux integers, according to equation (3.19)
of GKP. So the criterion for non-trivial warping is $e^{-4u}e^{-4A_{min}}>>1$.
Then we have for these KK modes which are localized down the throat\[
M_{KK}^{2}\sim\frac{1}{2\pi\alpha'}e^{4A_{min}}e^{-4u}.\]
Similarly down this throat the effective string scale is warped down
giving,\[
M_{s}^{2}\sim\frac{1}{2\pi\alpha'}e^{2A_{min}}e^{-4u}e^{\phi/2}.\]

\begin{equation}
\frac{M_{KK}^{2}}{M_{s}^{2}}\sim e^{-\phi/2}e^{2A_{min}}.\label{kkstring}\end{equation}
So the warped KK scale is below the warped string scale at least if
the string coupling is not too small. 

Now we may compute the ratios of the effective tension of the domain
wall wrapping a three cycle at the end of the throat to the warped
string scale, the Planck scale and the warped KK scale:\begin{eqnarray}
\frac{T_{dw}}{M_{s}^{3}} & \sim & e^{-\phi/4}e^{-3A_{min}}|z|\sim e^{-\phi/4}\label{dwstring}\\
\frac{T_{dw}}{M_{P}^{3}} & \sim & e^{-6u+\phi/2}|z|\sim e^{-6u+\phi/2}e^{3A_{min}}\label{dwplanck}\\
\frac{T_{dw}}{M_{KK}^{3}} & \sim & e^{\phi/2}e^{-6A_{min}}|z|\sim e^{\phi/2}e^{-3A_{min}}\label{dwkk}\end{eqnarray}
In these expressions we've ignored factors of $O(1)$ and retained
only the modulus/dilaton/warp factor dependences and in the last estimate
we've used the value $z\sim e^{3A_{min}}$ given in equation (3.19)
of GKP \cite{Giddings:2001yu}. 

From (\ref{dwplanck}) we see that the tension is always below the
Planck scale for large volume compactifications as observed in \cite{Kachru:2002ns}.
But from (\ref{dwstring}) we see that $T_{dw}$ is above the string
scale regardless of the size of the cycle relative to the bulk %
\footnote{Of course we are assuming that $e^{\phi}$ is not greater than one
- which is a starting assumption of all these effective field theory
arguments.%
}. The same is true of the ratio to the KK scale. In other words the
spontaneous nucleation of these domain walls (and hence bubbles of
different flux vacua) cannot be described within effective field theory.
To put it another way, in regions way down some KS throat, where massive
modes are likely to be localized, with $e^{A_{min}}<<1$, the domain
wall tension is above the local string scale and way above the local
KK scale.

\section{The BTBP process in IIB}

In the Brown Teitelboim process (adapted to a multi-flux/brane situation
by Bousso and Polchinski) it was assumed that the effective cosmological
constant (CC) is given by a formula of the form \[
\lambda=\lambda_{0}+\frac{1}{2}\sum_{i=1}^{J}n_{i}^{2}q_{i}^{2}.\]
Here the first term is the CC coming from all other sources (both
classical and quantum) and is assumed to be negative. The second term
is the explicit contribution of the fluxes with $\{ n_{i}\}$ being
a set of integers characterizing the flux configuration and $\{ q_{i}\}$
being the set of charges associated with branes wrapping any one of
the $J$ cycles in the compact manifold. In the original BT process
there was only one charge and then in order to obtain an effective
CC of $O(10^{-120})$ in Planck units it was necessary to have a charge
of the same order. However as Bousso and Polchinski showed if we have
many charges (as would be the case typically in string theory compactified
on a manifold with many cycles say $O(100)$) the discretuum of values
of the cosmological constant would be sufficiently dense, so that
even with relatively large charges (say $O(1/10)$) there would be
some which are within the acceptable range.

Within a semi-classical approximation BT had derived (in a manner
similar to that in the seminal work of Coleman and De Luccia \cite{Coleman:1980aw})
the probability of brane nucleation. It was argued that in 4-space
the effect of brane nucleation was to create a bubble of lower (or
higher) values of the cosmological constant separated by a domain
wall consisting of a closed brane configuration. Also while transitions
to lower values will take place at some finite rate, upward transitions
could not happen for non-compact 3-space, while if space was compact
they could take place but at some suppressed rate. Furthermore transitions
from a positive to a negative value of the CC were also suppressed.

However the BP argument was made in a context where the moduli (in
particular the sizes of the cycles and hence the values of the charges)
were not stabilized. So a proper discussion of this process requires
one to consider an explicit string theory model such as that of GKP-KKLT
\cite{Giddings:2001yu}\cite{Kachru:2003aw} in type IIB compactified
on a Calabi-Yau orientifold $X$. In order to cancel tadpoles one
introduces $D_{3}$ (and $D_{7}$) branes and turns on three form
fluxes. The fluxes generate a potential for the dilaton and complex
structure moduli ($z^{i}\, i,=1,\ldots,h_{21})$. By adding non-perturbative
effects a potential for the Kaehler moduli $T_{i}\, i=1,\ldots,h_{11}$
can also be generated. This potential is in the classic ${\mathcal{N}}=1$
SUGRA form and is given in terms of a Kaehler potential (assuming
for simplicity just one Kaehler modulus i.e. $h_{11}=1$)\begin{equation}
K=-\ln(S+\bar{S})-3\ln(T+\bar{T})-\ln[i\int_{X}\Omega\wedge\bar{\Omega}],\label{Kaehler}\end{equation}
and a superpotential

\begin{equation}
W=\int_{X}G_{3}\wedge\Omega+Ce^{-aT}.\label{W}\end{equation}
In the above $S=e^{-\phi}+iC_{0}$ and $T=e^{4u}+ib$ with $C_{0}$
being the RR zero-form and $b$ is an axion related to the RR four-form
field of IIB string theory. We also set $2\pi\alpha'=1$ in the following.
Also $\Omega$ is the holomorphic three form on $X$ and \[
G_{3}=F_{3}+iSH_{3}\]
 with $F_{3}(H_{3})$ being RR (NSNS) three form fluxes threading
some three cycles in $X$. As is well-known (see for example \cite{Grana:2005jc}
for a review) these fluxes are quantized and by expanding in a basis
of ($A$ and $B$) three cycles the first (flux) term of (\ref{W})
can be written as\begin{equation}
W_{flux}\equiv\int_{X}G_{3}\wedge\Omega=(2\pi)\sum_{i}[(n_{A}^{i}+iSm_{A}^{i}){\mathcal{G}}_{i}(z)-(n_{i}^{B}+iSm_{i}^{B})z^{i}].\label{fluxW}\end{equation}
Here the $z^{i}$ are projective coordinates (defined by the periods
of $\Omega$) on the complex structure moduli space, ${\mathcal{G}}{}_{i}=\partial{\mathcal{G}}/\partial z^{i}$
with ${\mathcal{G}}$ being a homogeneous function of degree two in
the $z^{i}$ and the n's and m's are integers.

An elementary brane nucleation transition (of the sort that was considered
by BT) would correspond for example to a shift of the form \[
n_{j}^{B}\rightarrow n_{j}^{B}\pm1,\]
for one flux integer with all others unchanged %
\footnote{The Bianchi identity for the five form flux would imply that there
is a corresponding change in the effective number of D3 branes. %
}. This would lead to a change of the superpotential\begin{equation}
\Delta W=\pm2\pi z^{j}.\label{deltaW}\end{equation}

To find the change in the potential and the corresponding change in
the stabilized values of the moduli is not easy in general. However
we may estimate these values by adopting the two stage procedure of
KKLT. So we assume that it is possible to integrate out the complex
structure moduli and the axi-dilaton before adding the NP term giving
a constant flux superpotential (for each choice of fluxes) and then
consider a simple theory of the one remaining modulus $T$. This means
taking $K=-3\ln(T+\bar{T})+O(1)$ and $W=W_{0}+Ce^{-aT}$ giving a
potential (setting for simplicity the imaginary part $T_{I}=0$) \begin{equation}
V=v\frac{aCe^{-aT_{R}}}{2T_{R}^{2}}[W_{0}+Ce^{-aT_{R}}(1+\frac{1}{3}aT_{R})]\label{KKLTpot}\end{equation}
with $v$ an O(1) constant. This potential has a supersymmetric AdS
minimum determined by $D_{T}W$=0, i.e. \begin{equation}
W_{0}=-(1+\frac{2T_{R}a}{3})Ce^{-aT_{R}}.\label{minKKLT}\end{equation}
KKLT go on to lift this minimum to get a dS solution by a rather ad
hoc procedure. For our purposes we do not need this - in fact before
we take into account the effects of integrating down from just below
the string scale all the way to the Hubble scale (where the CC is
determined) all that one needs to ensure is that the CC produced from
a string theory argument is of the order of the TeV scale, since one
expects an effect of at least this order from supersymmery breaking,
the standard model phase transition and associated quantum fluctuations.
Thus we will take the above minimum as representing the one that we
find ourselves in, modulo the effects just mentioned.

The question then is what was the prior state which gave rise to this
flux configuration. First we note that for the scheme to be valid
(i.e. $T_{R}>1$, and $aT_{R}>1$ as pointed out by KKLT) we should
have $W_{0}\ll1$. Indeed to have a CC at the standard model scale
we need (calling the final (initial) value of the CC $V_{0}^{f}(V_{0}^{i})$)\begin{equation}
V_{0}^{f}\sim-\frac{|W_{0}^{f}|^{2}}{T_{R}^{3}}\ll1.\label{finalmin}\end{equation}

Let us now estimate the CC of the immediate prior state to this -
in other words the potential minimum of the state from which this
universe was created by bubble nucleation. The change in the superpotential
is given in (\ref{deltaW}) and since this is much greater than $W_{0}$,
the CC of the state from which the final state was nucleated had a
CC of the order of\begin{equation}
|V_{0}^{i}|\sim\frac{z^{2}}{T_{R}^{3}}\sim\frac{e^{6A_{min}}}{T_{R}^{3}}.\label{initialmin}\end{equation}
 In the last step we have assumed that the three cycle which is wrapped
by the nucleated five-brane is the three cycle in the KS throat with
$z$ the corresponding complex structure modulus which was estimated
in equation (3.19) of GKP \cite{Giddings:2001yu}. Let us now measure
this against the string scale using (\ref{stringtension}) .\begin{equation}
\frac{V_{0}^{i}}{M_{s}^{4}}\sim e^{2A_{min}}e^{-\phi}e^{-4u}\sim\frac{1}{g_{s}}e^{-4\pi K/3Mg_{s}}e^{-4u}<<1,\label{Vi/Ms}\end{equation}
 where $K,M$ are integers characterizing the fluxes through the relevant
pair of dual three cycles. We are assuming of course that the values
of the moduli (in particular $T_{R}$ in the initial and final state
are not parametrically different, consistent with \eqref{deltaW}
and \eqref{minKKLT}.

Thus we have shown that although one might expect that the generic
initial state has a CC that is around the string scale, so there is
no possibility of a supergravity description of the initial state,
for strongly warped large volume configurations, it is possible to
have such a description %
\footnote{In the extreme case of the cycle shrinking almost to zero this would
be equivalent to the version of the BT process of \cite{Feng:2000if}.%
}. However as we argued at the end of the previous section (see (\ref{dwkk})(\ref{dwstring})
) even for branes wrapping the shrinking three cycle down the KS throat,
the domain wall tension is larger than the string scale and much larger
than the KK scale!

Thus it seems that even though there are flux configurations with
a strongly warped throat that allow a supergravity description of
both the final state and the initial state - the brane nucleation
process requires a brane whose effective tension is well above the
(local) string scale. In contrast to working in string perturbation
theory with a static background of D-branes (which have tensions that
are larger than the string scale) here we have dynamical processes
that are taking place at or above the string scale. The brane \textit{nucleation}
involves a redistribution of energy density at scales which are greater
than or equal to the string scale, and therefore one cannot really
expect there to be a low-energy effective field theory description
of this. One should expect that at the same time as these branes are
nucleated, stringy states could also be created. Even if it is the
case that Coleman-De Luccia type calculations \cite{Coleman:1980aw}
are still valid for estimating the transition probabilities, the end
point of the transition is unlikely to have a semi-classical description.
As pointed out in \cite{deAlwis:2006cb} there is also a problem of
constructing a background independent effective supergravity action
that describes such processes. Of course this does not mean that there
are no such transitions in string theory. It simply casts doubt on
the existence of a supergravity description of the dynamics.

\section{Acknowledgments}

I wish to thank Cliff Burgess and Kerim Suruliz for stimulating discussions
on warped compactifications and Fernando Quevedo for useful correspondence.
In addition I wish to thank the members of the landscape discussion
group (especially Tom Banks, Steve Giddings and Anshuman Maharanna)
at the KITP workshop on String Phenomenology, for useful comments.
I also wish to acknowledge the hospitality of KITP while this work
was being revised. This research is supported by DOE grant No. DE-FG02-91-ER-40672
and the Perimeter Institute.

\bibliographystyle{apsrev}
\bibliography{myrefs}

\begin{thebibliography}{24}
\expandafter\ifx\csname natexlab\endcsname\relax\def\natexlab#1{#1}\fi
\expandafter\ifx\csname bibnamefont\endcsname\relax
  \def\bibnamefont#1{#1}\fi
\expandafter\ifx\csname bibfnamefont\endcsname\relax
  \def\bibfnamefont#1{#1}\fi
\expandafter\ifx\csname citenamefont\endcsname\relax
  \def\citenamefont#1{#1}\fi
\expandafter\ifx\csname url\endcsname\relax
  \def\url#1{\texttt{#1}}\fi
\expandafter\ifx\csname urlprefix\endcsname\relax\def\urlprefix{URL }\fi
\providecommand{\bibinfo}[2]{#2}
\providecommand{\eprint}[2][]{\url{#2}}

\bibitem[{\citenamefont{Hartle and Hawking}(1983)}]{Hartle:1983ai}
\bibinfo{author}{\bibfnamefont{J.~B.} \bibnamefont{Hartle}} \bibnamefont{and}
  \bibinfo{author}{\bibfnamefont{S.~W.} \bibnamefont{Hawking}},
  \bibinfo{journal}{Phys. Rev.} \textbf{\bibinfo{volume}{D28}},
  \bibinfo{pages}{2960} (\bibinfo{year}{1983}).

\bibitem[{\citenamefont{Vilenkin}(1983)}]{Vilenkin:1983xq}
\bibinfo{author}{\bibfnamefont{A.}~\bibnamefont{Vilenkin}},
  \bibinfo{journal}{Phys. Rev.} \textbf{\bibinfo{volume}{D27}},
  \bibinfo{pages}{2848} (\bibinfo{year}{1983}).

\bibitem[{\citenamefont{Linde}(1984)}]{Linde:1983cm}
\bibinfo{author}{\bibfnamefont{A.~D.} \bibnamefont{Linde}},
  \bibinfo{journal}{Sov. Phys. JETP} \textbf{\bibinfo{volume}{60}},
  \bibinfo{pages}{211} (\bibinfo{year}{1984}).

\bibitem[{\citenamefont{Brown and Teitelboim}(1988)}]{Brown:1988kg}
\bibinfo{author}{\bibfnamefont{J.~D.} \bibnamefont{Brown}} \bibnamefont{and}
  \bibinfo{author}{\bibfnamefont{C.}~\bibnamefont{Teitelboim}},
  \bibinfo{journal}{Nucl. Phys.} \textbf{\bibinfo{volume}{B297}},
  \bibinfo{pages}{787} (\bibinfo{year}{1988}).

\bibitem[{\citenamefont{Bousso and Polchinski}(2000)}]{Bousso:2000xa}
\bibinfo{author}{\bibfnamefont{R.}~\bibnamefont{Bousso}} \bibnamefont{and}
  \bibinfo{author}{\bibfnamefont{J.}~\bibnamefont{Polchinski}},
  \bibinfo{journal}{JHEP} \textbf{\bibinfo{volume}{06}}, \bibinfo{pages}{006}
  (\bibinfo{year}{2000}), \eprint{hep-th/0004134}.

\bibitem[{\citenamefont{Brustein and de~Alwis}(2006)}]{Brustein:2005yn}
\bibinfo{author}{\bibfnamefont{R.}~\bibnamefont{Brustein}} \bibnamefont{and}
  \bibinfo{author}{\bibfnamefont{S.~P.} \bibnamefont{de~Alwis}},
  \bibinfo{journal}{Phys. Rev.} \textbf{\bibinfo{volume}{D73}},
  \bibinfo{pages}{046009} (\bibinfo{year}{2006}), \eprint{hep-th/0511093}.

\bibitem[{\citenamefont{Kachru et~al.}(2003{\natexlab{a}})\citenamefont{Kachru,
  Kallosh, Linde, and Trivedi}}]{Kachru:2003aw}
\bibinfo{author}{\bibfnamefont{S.}~\bibnamefont{Kachru}},
  \bibinfo{author}{\bibfnamefont{R.}~\bibnamefont{Kallosh}},
  \bibinfo{author}{\bibfnamefont{A.}~\bibnamefont{Linde}}, \bibnamefont{and}
  \bibinfo{author}{\bibfnamefont{S.~P.} \bibnamefont{Trivedi}},
  \bibinfo{journal}{Phys. Rev.} \textbf{\bibinfo{volume}{D68}},
  \bibinfo{pages}{046005} (\bibinfo{year}{2003}{\natexlab{a}}),
  \eprint{hep-th/0301240}.

\bibitem[{\citenamefont{Giddings et~al.}(2002)\citenamefont{Giddings, Kachru,
  and Polchinski}}]{Giddings:2001yu}
\bibinfo{author}{\bibfnamefont{S.~B.} \bibnamefont{Giddings}},
  \bibinfo{author}{\bibfnamefont{S.}~\bibnamefont{Kachru}}, \bibnamefont{and}
  \bibinfo{author}{\bibfnamefont{J.}~\bibnamefont{Polchinski}},
  \bibinfo{journal}{Phys. Rev.} \textbf{\bibinfo{volume}{D66}},
  \bibinfo{pages}{106006} (\bibinfo{year}{2002}), \eprint{hep-th/0105097}.

\bibitem[{\citenamefont{Randall and Sundrum}(1999)}]{Randall:1999ee}
\bibinfo{author}{\bibfnamefont{L.}~\bibnamefont{Randall}} \bibnamefont{and}
  \bibinfo{author}{\bibfnamefont{R.}~\bibnamefont{Sundrum}},
  \bibinfo{journal}{Phys. Rev. Lett.} \textbf{\bibinfo{volume}{83}},
  \bibinfo{pages}{3370} (\bibinfo{year}{1999}), \eprint{hep-ph/9905221}.

\bibitem[{\citenamefont{DeWolfe and Giddings}(2003)}]{DeWolfe:2002nn}
\bibinfo{author}{\bibfnamefont{O.}~\bibnamefont{DeWolfe}} \bibnamefont{and}
  \bibinfo{author}{\bibfnamefont{S.~B.} \bibnamefont{Giddings}},
  \bibinfo{journal}{Phys. Rev.} \textbf{\bibinfo{volume}{D67}},
  \bibinfo{pages}{066008} (\bibinfo{year}{2003}), \eprint{hep-th/0208123}.

\bibitem[{\citenamefont{Goldberger and Wise}(1999)}]{Goldberger:1999uk}
\bibinfo{author}{\bibfnamefont{W.~D.} \bibnamefont{Goldberger}}
  \bibnamefont{and} \bibinfo{author}{\bibfnamefont{M.~B.} \bibnamefont{Wise}},
  \bibinfo{journal}{Phys. Rev. Lett.} \textbf{\bibinfo{volume}{83}},
  \bibinfo{pages}{4922} (\bibinfo{year}{1999}), \eprint{hep-ph/9907447}.

\bibitem[{\citenamefont{Frey and Maharana}(2006)}]{Frey:2006wv}
\bibinfo{author}{\bibfnamefont{A.~R.} \bibnamefont{Frey}} \bibnamefont{and}
  \bibinfo{author}{\bibfnamefont{A.}~\bibnamefont{Maharana}}
  (\bibinfo{year}{2006}), \eprint{hep-th/0603233}.

\bibitem[{\citenamefont{Giddings and Maharana}(2006)}]{Giddings:2005ff}
\bibinfo{author}{\bibfnamefont{S.~B.} \bibnamefont{Giddings}} \bibnamefont{and}
  \bibinfo{author}{\bibfnamefont{A.}~\bibnamefont{Maharana}},
  \bibinfo{journal}{Phys. Rev.} \textbf{\bibinfo{volume}{D73}},
  \bibinfo{pages}{126003} (\bibinfo{year}{2006}), \eprint{hep-th/0507158}.

\bibitem[{\citenamefont{Kachru et~al.}(2003{\natexlab{b}})\citenamefont{Kachru,
  Liu, Schulz, and Trivedi}}]{Kachru:2002ns}
\bibinfo{author}{\bibfnamefont{S.}~\bibnamefont{Kachru}},
  \bibinfo{author}{\bibfnamefont{X.}~\bibnamefont{Liu}},
  \bibinfo{author}{\bibfnamefont{M.~B.} \bibnamefont{Schulz}},
  \bibnamefont{and} \bibinfo{author}{\bibfnamefont{S.~P.}
  \bibnamefont{Trivedi}}, \bibinfo{journal}{JHEP}
  \textbf{\bibinfo{volume}{05}}, \bibinfo{pages}{014}
  (\bibinfo{year}{2003}{\natexlab{b}}), \eprint{hep-th/0205108}.

\bibitem[{\citenamefont{Coleman and De~Luccia}(1980)}]{Coleman:1980aw}
\bibinfo{author}{\bibfnamefont{S.~R.} \bibnamefont{Coleman}} \bibnamefont{and}
  \bibinfo{author}{\bibfnamefont{F.}~\bibnamefont{De~Luccia}},
  \bibinfo{journal}{Phys. Rev.} \textbf{\bibinfo{volume}{D21}},
  \bibinfo{pages}{3305} (\bibinfo{year}{1980}).

\bibitem[{\citenamefont{Grana}(2006)}]{Grana:2005jc}
\bibinfo{author}{\bibfnamefont{M.}~\bibnamefont{Grana}},
  \bibinfo{journal}{Phys. Rept.} \textbf{\bibinfo{volume}{423}},
  \bibinfo{pages}{91} (\bibinfo{year}{2006}), \eprint{hep-th/0509003}.

\bibitem[{\citenamefont{de~Alwis}(2006)}]{deAlwis:2006cb}
\bibinfo{author}{\bibfnamefont{S.~P.} \bibnamefont{de~Alwis}}
  (\bibinfo{year}{2006}), \eprint{hep-th/0605184}.

\bibitem[{\citenamefont{Sarangi and Tye}(2005)}]{Sarangi:2005cs}
\bibinfo{author}{\bibfnamefont{S.}~\bibnamefont{Sarangi}} \bibnamefont{and}
  \bibinfo{author}{\bibfnamefont{S.~H.~H.} \bibnamefont{Tye}}
  (\bibinfo{year}{2005}), \eprint{hep-th/0505104}.

\bibitem[{\citenamefont{Kachru et~al.}(2002)\citenamefont{Kachru, Pearson, and
  Verlinde}}]{Kachru:2002gs}
\bibinfo{author}{\bibfnamefont{S.}~\bibnamefont{Kachru}},
  \bibinfo{author}{\bibfnamefont{J.}~\bibnamefont{Pearson}}, \bibnamefont{and}
  \bibinfo{author}{\bibfnamefont{H.~L.} \bibnamefont{Verlinde}},
  \bibinfo{journal}{JHEP} \textbf{\bibinfo{volume}{06}}, \bibinfo{pages}{021}
  (\bibinfo{year}{2002}), \eprint{hep-th/0112197}.

\bibitem[{\citenamefont{Frey et~al.}(2003)\citenamefont{Frey, Lippert, and
  Williams}}]{Frey:2003dm}
\bibinfo{author}{\bibfnamefont{A.~R.} \bibnamefont{Frey}},
  \bibinfo{author}{\bibfnamefont{M.}~\bibnamefont{Lippert}}, \bibnamefont{and}
  \bibinfo{author}{\bibfnamefont{B.}~\bibnamefont{Williams}},
  \bibinfo{journal}{Phys. Rev.} \textbf{\bibinfo{volume}{D68}},
  \bibinfo{pages}{046008} (\bibinfo{year}{2003}), \eprint{hep-th/0305018}.

\bibitem[{\citenamefont{Pilo et~al.}(2004)\citenamefont{Pilo, Riotto, and
  Zaffaroni}}]{Pilo:2004mg}
\bibinfo{author}{\bibfnamefont{L.}~\bibnamefont{Pilo}},
  \bibinfo{author}{\bibfnamefont{A.}~\bibnamefont{Riotto}}, \bibnamefont{and}
  \bibinfo{author}{\bibfnamefont{A.}~\bibnamefont{Zaffaroni}},
  \bibinfo{journal}{JHEP} \textbf{\bibinfo{volume}{07}}, \bibinfo{pages}{052}
  (\bibinfo{year}{2004}), \eprint{hep-th/0401004}.

\bibitem[{\citenamefont{de~Alwis}(2003)}]{deAlwis:2003sn}
\bibinfo{author}{\bibfnamefont{S.~P.} \bibnamefont{de~Alwis}},
  \bibinfo{journal}{Phys. Rev.} \textbf{\bibinfo{volume}{D68}},
  \bibinfo{pages}{126001} (\bibinfo{year}{2003}), \eprint{hep-th/0307084}.

\bibitem[{\citenamefont{de~Alwis}(2004)}]{deAlwis:2004qh}
\bibinfo{author}{\bibfnamefont{S.~P.} \bibnamefont{de~Alwis}},
  \bibinfo{journal}{Phys. Lett.} \textbf{\bibinfo{volume}{B603}},
  \bibinfo{pages}{230} (\bibinfo{year}{2004}), \eprint{hep-th/0407126}.

\bibitem[{\citenamefont{Feng et~al.}(2001)\citenamefont{Feng, March-Russell,
  Sethi, and Wilczek}}]{Feng:2000if}
\bibinfo{author}{\bibfnamefont{J.~L.} \bibnamefont{Feng}},
  \bibinfo{author}{\bibfnamefont{J.}~\bibnamefont{March-Russell}},
  \bibinfo{author}{\bibfnamefont{S.}~\bibnamefont{Sethi}}, \bibnamefont{and}
  \bibinfo{author}{\bibfnamefont{F.}~\bibnamefont{Wilczek}},
  \bibinfo{journal}{Nucl. Phys.} \textbf{\bibinfo{volume}{B602}},
  \bibinfo{pages}{307} (\bibinfo{year}{2001}), \eprint{hep-th/0005276}.

\end{thebibliography}

\end{document}